\documentclass[twocolumn,prl,aps,superscriptaddress,footnote,amsfonts,showpacs]{revtex4-1}
\usepackage{amsmath}
\usepackage{amssymb}
\usepackage{bm}
\usepackage{epsfig}
\usepackage{color}
\usepackage{wrapfig}

\begin{document}
\title{Towards an explanation of transverse single-spin asymmetries in \\ 
proton-proton collisions:~$\!$the role of fragmentation in collinear factorization}

\author{Koichi Kanazawa}
\affiliation{Graduate School of Science and Technology,
                 Niigata University, Ikarashi,
                 Niigata 950-2181, Japan}
\affiliation{Department of Physics, Barton Hall,
                 Temple University, Philadelphia,
                 Pennsylvania 19122, USA}

\author{Yuji Koike}
\affiliation{Department of Physics,
                 Niigata University, Ikarashi,
                 Niigata 950-2181, Japan}

\author{Andreas Metz}
\affiliation{Department of Physics, Barton Hall,
                 Temple University, Philadelphia,
                 Pennsylvania 19122, USA}

\author{Daniel Pitonyak}
\affiliation{RIKEN BNL Research Center,
                 Brookhaven National Laboratory,
                 Upton, New York 11973, USA}

\begin{abstract}
We study the transverse single-spin asymmetry for single-hadron production in proton-proton collisions within the framework of collinear twist-3 factorization in Quantum Chromodynamics. 
By taking into account the contribution due to parton fragmentation we obtain a very good description of all high transverse-momentum data for neutral and charged pion production from the Relativistic Heavy Ion Collider.
Our study may provide the crucial step towards a final solution to the longstanding problem of what causes transverse single-spin asymmetries in hadronic collisions within Quantum Chromodynamics.
We show for the first time that it is possible to simultaneously describe spin/azimuthal asymmetries in proton-proton collisions, semi-inclusive deep-inelastic scattering, and electron-positron annihilation by using collinear twist-3 factorization in the first process along with transverse momentum dependent functions extracted from the latter two reactions.
\end{abstract}


\maketitle

%
%
{\it Introduction}
$\phantom{A}$The field of transverse single-spin asymmetries (SSAs) in hard semi-inclusive processes began some four decades ago with the observation of the large transverse polarization (up to about 30\,\%) of neutral $\Lambda$-hyperons in the process $p \, Be \to \Lambda^{\uparrow} X$ at FermiLab~\cite{Bunce:1976yb}. 
People noticed early on that the na\"{i}ve collinear parton model cannot generate such large effects~\cite{Kane:1978nd}.
It was then pointed out that SSAs for single-particle production in hadronic collisions are genuine twist-3 observables for which, in particular, collinear 3-parton correlations have to be taken into account in order to have a proper description within Quantum Chromodynamics (QCD)~\cite{Efremov:1981sh}.
This formalism later on was worked out in more detail and also successfully applied to SSAs in processes like hadron production in proton-proton collisions, $p^{\uparrow} p \to h X$ --- see, e.g., Refs.~\cite{Qiu:1991pp,Qiu:1998ia,Eguchi:2006qz,Kouvaris:2006zy,Koike:2007rq,Koike:2009ge,Kanazawa:2010au}.
Here we focus on SSAs in such reactions, which were extensively investigated in fixed target and in collider experiments.

Let us now look at the generic structure of the spin-dependent cross section for $A(P,\,\vec{S}_{\perp}) + B(P') \rightarrow C(P_{h}) + X$,
where the 4-momenta and polarizations of the incoming protons $A$, $B$ and outgoing hadron $C$ are specified.
In twist-3 collinear QCD factorization one has
\begin{align} 
d\sigma(\vec{S}_{\perp}) &= \,H\otimes f_{a/A(3)}\otimes f_{b/B(2)}\otimes D_{C/c(2)} \nonumber \\*
&+ \,H'\otimes f_{a/A(2)}\otimes f_{b/B(3)}\otimes D_{C/c(2)} \nonumber \\
&+ \,H''\otimes f_{a/A(2)}\otimes f_{b/B(2)}\otimes D_{C/c(3)}\,,
\label{e:sigma_generic}
\end{align} 
with $f_{a/A(t)}$ ($f_{b/B(t)}$) indicating the distribution function associated with parton $a$ ($b$) in proton $A$ ($B$), while $D_{C/c(t)}$ represents the fragmentation function associated with hadron $C$ in parton $c$.  
The twist of the functions is denoted by $t$.  
The hard factors corresponding to each term are given by $H$, $H'$, and $H''$, and the symbol $\otimes$ represents convolutions in the appropriate momentum fractions.  
In Eq.~(\ref{e:sigma_generic}) a sum over partonic channels and parton flavors in each channel is understood.

The first term in~(\ref{e:sigma_generic}) has already been studied in quite some detail in the literature~\cite{Qiu:1998ia, Kouvaris:2006zy, Koike:2007rq, Koike:2009ge,Kanazawa:2010au,Kang:2011hk,Beppu:2013uda}.  
It contains both quark-gluon-quark correlations and tri-gluon correlations in the polarized proton, where for the former one needs to distinguish between contributions from so-called soft gluon poles (SGPs) and soft fermion poles (SFPs).
The second term in~(\ref{e:sigma_generic}), arising from twist-3 effects in the unpolarized proton, was shown to be small~\cite{Kanazawa:2000hz}.
Only recently a complete analytical result was obtained for the third term in~(\ref{e:sigma_generic}), which describes the twist-3 contribution due to parton fragmentation~\cite{Metz:2012ct}.

For quite some time many in the community believed that the first term in~(\ref{e:sigma_generic}) dominates the transverse SSA in~$p^{\uparrow} p \to h X$ (typically denoted by $A_N$) for the production of light hadrons (see, e.g., Refs.~\cite{Qiu:1998ia,Kouvaris:2006zy,Kanazawa:2010au}), where the SGP contribution is generally considered the most important part. 
Note that the SGP contribution to $A_N$ is determined by the Qiu-Sterman function $T_F$~\cite{Qiu:1991pp,Qiu:1998ia}, which can be related to the transverse-momentum dependent (TMD) Sivers parton density $f_{1T}^{\perp}$~\cite{Sivers:1989cc,Boer:1997nt}.
For a given quark flavor $q$, these entities satisfy~\cite{Boer:2003cm}
\begin{equation}
T_F^q(x,x) = -\int d^2\vec{p}_{\perp} \, \frac{\vec{p}_{\perp}^{\,2}}{M} \, f_{1T}^{\perp q}(x,\vec{p}_{\perp}^{\,2})\big|_{\textrm{SIDIS}} \,,
\label{e:TF_Siv}
\end{equation}
where $M$ is the nucleon mass.
Because of the relation in~(\ref{e:TF_Siv}), one can extract $T_F$ from data on either $A_N$ or on the Sivers transverse SSA in semi-inclusive deep-inelastic scattering (SIDIS) $A_{\textrm{SIDIS}}^{\textrm{Siv}}$.  It therefore came as a major surprise when an attempt failed to simultaneously explain both $A_N$ and $A_{\textrm{SIDIS}}^{\textrm{Siv}}$~\cite{Kang:2011hk}.  
The striking result pointed out in Ref.~\cite{Kang:2011hk} was that the two extractions for $T_F$ differ in sign.
This ``sign-mismatch'' puzzle could not be resolved by more flexible parameterizations of $f_{1T}^{\perp}$~\cite{Kang:2012xf}. 
Also tri-gluon correlations are unlikely to fix this issue~\cite{Beppu:2013uda}, while SFPs may play some role~\cite{Koike:2009ge}.

At this point one may start to question the dominance of the first term in~(\ref{e:sigma_generic}).
In fact, data on the transverse SSA in inclusive DIS~\cite{Airapetian:2009ab,Katich:2013atq} seem to support this point of view, i.e., that the first term in~(\ref{e:sigma_generic}) is not the main cause of $A_N$~\cite{Metz:2012ui}.
Therefore, in the present work we study the potential role of the twist-3 fragmentation part of~(\ref{e:sigma_generic}).
After fixing the SGP contribution to $A_N$ through the Sivers function extracted from data on $A_{\textrm{SIDIS}}^{\textrm{Siv}}$~\cite{Anselmino:2008sga,Anselmino:2013rya} and the relation in~(\ref{e:TF_Siv}), we obtain a very good fit to all high transverse-momentum forward-region pion data for $A_N$ from the Relativistic Heavy Ion Collider (RHIC).
As explained below in more detail, our analysis shows for the first time that one can simultaneously describe $A_N$ using collinear factorization, $A_{\textrm{SIDIS}}^{\textrm{Siv}}$, the Collins transverse SSA $A_{\textrm{SIDIS}}^{\textrm{Col}}$ in SIDIS, and $A_{e^+ e^-}^{\cos(2\phi)}$  that represents a particular azimuthal asymmetry in electron-positron annihilation into two hadrons, $e^+ e^- \to h_1 h_2 X$~\cite{Boer:1997mf}.

%
%
\vspace{0.1cm}
{\it Fragmentation contribution to $A_N$}$
\phantom{A}$The fragmentation contribution to the cross section in (1) reads~\cite{Metz:2012ct}
\begin{eqnarray}
\frac{P_{h}^{0}d\sigma(\vec{S}_{\perp})} {d^{3}\vec{P}_{h}} \! &=& \! -\frac{2\alpha_{s}^{2}M_{h}} {S} \epsilon_{\perp,\alpha\beta}\,S_{\perp}^{\alpha}P_{h\perp}^{\beta} \sum_{i}\sum_{a,b,c}\int_{z_{min}}^{1}\frac{dz} {z^{3}}
\nonumber \\
&& \hspace{-1.0cm} \times \int_{x'_{min}}^{1}\frac{d x'} {x'} \frac{1} {x}\,\frac{1} {x' S+T/z}\,\frac{1} {-x'\hat{t}-x\hat{u}} \,h_{1}^{a}(x)\,f_{1}^{b}(x')
\nonumber \\
&&\hspace{-1.0cm}\times\,\bigg\{ \bigg[\hat{H}^{C/c}(z)-z\frac{d\hat{H}^{C/c}(z)} {dz}\bigg]\,S_{\hat{H}}^{i} + \frac{1} {z} H^{C/c}(z)\, S_{H}^{i}\nonumber \\
&& \hspace{-1.0cm} + \,2z^2\int_{z}^{\infty} \frac{dz_1} {z_1^2} \frac{1} {\frac{1} {z}-\frac{1} {z_{1}}} \, \hat{H}_{FU}^{C/c,\Im}(z,z_{1})\,\frac{1} {\xi} \,S_{\hat{H}_{FU}}^{i}\bigg\}\,, \phantom{AA}
\label{e:sigma_frag}
\end{eqnarray}
where $i$ denotes the channel, $x=-x'(U/z)/(x'S+T/z)$, $x'_{min}=-(T/z)/(U/z+S)$, $z_{min}=-(T+U)/S$, and $\xi = (1-z/z_1)$.
Here we used the Mandelstam variables $S = (P+P')^{2}$, $T = (P-P_h)^{2}$, and $U = (P'-P_h)^{2}$, which on the partonic level give $\hat{s} = xx' S$, $\hat{t} = xT/z$, and $\hat{u} = x' U/z$. 
Oftentimes one also uses $x_F = 2P_{hz}/\sqrt{S}$, where $P_{hz}$ is the longitudinal momentum of the hadron, as well as the pseudo-rapidity $\eta = - \ln \tan(\theta/2)$, where $\theta$ is the scattering angle.  The variables $x_F,\,\eta$ are further related by $x_F = 2P_{h\perp}\sinh(\eta)/\sqrt{S}$, where $P_{h\perp}$ is the transverse momentum of the hadron. 
The non-perturbative parts in (\ref{e:sigma_frag}) are the transversity distribution $h_1$, the unpolarized parton density $f_1$, and the three (twist-3) fragmentation functions (FFs) $\hat{H}$, $H$, and $\hat{H}_{FU}^{\Im}$, with the last one parameterizing the imaginary part  of a 3-parton correlator.
The definition of those functions and the results for the hard scattering coefficients $S^{i}$ can be found in~\cite{Metz:2012ct}.
(An alternative notation of the relevant FFs is given in Ref.~\cite{Kanazawa:2013uia}, where twist-3 effects in SIDIS were computed.)
We note that the so-called derivative term in~(\ref{e:sigma_frag}), associated with $d\hat{H}/dz$, was first obtained in~\cite{Kang:2010zzb}.

The function $\hat{H}$ is related to the TMD Collins function $H_1^\perp$~\cite{Collins:1992kk} according to~\cite{Kang:2010zzb,Metz:2012ct}
\begin{equation}
\hat{H}^{h/q}(z) = z^2\int d^2 \vec{k}_{\perp} \, \frac{\vec{k}_{\perp}^{\,2}}{2 M_h^2} \, 
H_{1}^{\perp \hspace{0.025cm}h/q}(z,z^2\vec{k}_{\perp}^{\,2})\,.
 \label{e:Hhat_Col}
\end{equation}
This relation can be considered the fragmentation counterpart of Eq.~(\ref{e:TF_Siv}).
Exploiting the universality of the Collins function~\cite{Metz:2002iz}, one can simultaneously extract $H_1^\perp$ and $h_1$ from data on $A_{\textrm{SIDIS}}^{\textrm{Col}}$~\cite{Airapetian:2010ds,Adolph:2012sn} and data on $A_{e^+ e^-}^{\cos(2\phi)}$~\cite{Seidl:2008xc,TheBABAR:2013yha} (see~\cite{Anselmino:2013vqa} and references therein).
Below we utilize such information for $H_1^\perp$ and $h_1$ when describing $A_N$.
The FFs in~(\ref{e:sigma_frag}) are related via~\cite{Metz:2012ct}
\begin{equation}
H^{h/q}(z) \! = \! - 2 z \hat{H}^{h/q}(z) 
           + 2 z^3 \!\! \int_{z}^{\infty} \frac{dz_1} {z_1^2} \!\! \frac{1} {\frac{1} {z}-\frac{1} {z_{1}}} \hat{H}_{FU}^{h/q,\Im}(z,z_{1}) \,,
\label{e:relation}
\end{equation}
implying that in the collinear twist-3 framework one has two independent FFs.
It is important to realize that this is different from the so-called TMD approach for $A_N$, where only $H_1^\perp$ enters the fragmentation piece~\cite{Anselmino:2012rq}.

%
%
\vspace{0.1cm}
{\it Phenomenology of $A_N$ for pion production}
$\phantom{A}$We consider $A_N$ for $p^{\uparrow} p \to \pi X$ in the forward region of the polarized proton, which has been studied by the STAR~\cite{Adams:2003fx,:2008qb,Adamczyk:2012xd}, BRAHMS~\cite{Lee:2007zzh,:2008mi} and PHENIX~\cite{Adare:2013ekj} collaborations at RHIC.
We mainly focus on data taken at $\sqrt{S} = 200 \, \textrm{GeV}$ for which typically $P_{h\perp} > 1 \, \textrm{GeV}$.
Throughout we use the GRV98 unpolarized parton distributions~\cite{Gluck:1998xa} and the DSS unpolarized FFs~\cite{deFlorian:2007aj}.
Note that the GRV98 parton distributions were also used in Refs.~\cite{Anselmino:2008sga,Anselmino:2013rya,Anselmino:2013vqa} for extracting the Sivers function and the transversity, which we take as input in our calculation.
The SGP contribution to~(\ref{e:sigma_generic}) is computed by fixing $T_F$ through Eq.~(\ref{e:TF_Siv}) with two different inputs for the Sivers function ---
SV1: $f_{1T}^\perp$ from Ref.~\cite{Anselmino:2008sga}, obtained from SIDIS data on $A_{\textrm{SIDIS}}^{\textrm{Siv}}$~\cite{Alekseev:2008aa,Airapetian:2009ae};
and SV2: $f_{1T}^\perp$ from Ref.~\cite{Anselmino:2013rya}, ``constructed" such that, in the TMD approach, the contribution of the Sivers effect to $A_N$ is maximized while maintaining a good description of $A_{\textrm{SIDIS}}^{\textrm{Siv}}$.
These two inputs for $f_{1T}^\perp$ are mainly distinct by their quite different large-$x$ behavior.
To compute the contribution in~(\ref{e:sigma_frag}) we take $h_1$ and $H_1^\perp$ (which fixes $\hat{H}$ through~(\ref{e:Hhat_Col})) from~\cite{Anselmino:2013vqa}.
For favored fragmentation into $\pi^+$ we make for $\hat{H}_{FU}^{\Im}$ the ansatz
\begin{eqnarray} 
\frac{\hat{H}_{FU}^{\pi^+/(u,\bar{d}) , \Im}(z,z_{1})} {D^{\pi^+ / (u,\bar{d})}(z) \, D^{\pi^+ / (u,\bar{d})}(z/z_1)} 
& = & \frac{N_{\textrm{fav}}}{2 I_{\textrm{fav}} J_{\textrm{fav}}} \, z^{\alpha_{\textrm{fav}}} (z/z_1)^{\alpha'_{\textrm{fav}}}
\nonumber \\
&& \hspace{-1.25cm} \times \, (1 - z)^{\beta_{\textrm{fav}}} \, (1 - z/z_{1})^{\beta'_{\textrm{fav}}} \,,
\label{e:par_fav}
\end{eqnarray}
with the parameters $N_{\textrm{fav}}$, $\alpha_{\textrm{fav}}$, $\alpha'_{\textrm{fav}}$, $\beta_{\textrm{fav}}$, $\beta'_{\textrm{fav}}$ and the unpolarized FF $D$.
Note that the allowed range for $z$ and $z/z_1$ is $[0,1]$~\cite{Meissner:2008yf} and that our ansatz satisfies the constraint $\hat{H}_{FU}(z,z) = 0$~\cite{Gamberg:2008yt,Meissner:2008yf}. 
With the use of DSS FFs~\cite{deFlorian:2007aj}, 
the factor $I_{\textrm{fav}}$ reads $I_{\textrm{fav}}\equiv I_{u+\bar{u}}-I_{\bar{u}}$
where $I_i$ ($i=u+\bar{u},\,\bar{u}$) is defined as 
\begin{eqnarray}
I_i& = & \frac{N_i (K_{1,\textrm{fav}} + \gamma_i K_{2,\textrm{fav}})}
{B[2 + \alpha_i, \beta_i + 1] + \gamma_i B[2 + \alpha_i, \beta_i + \delta_i + 1]} \,,
\nonumber \\[0.1cm]
&& \hspace{-0.5cm} \textrm{with} \;\,
K_{1,\textrm{fav}} = B[\alpha'_{\textrm{fav}} + \alpha_i + 1, \beta'_{\textrm{fav}} + 
\beta_i] \,, 
\\[0.1cm]
&& \hspace{0.35cm} 
K_{2,\textrm{fav}} = B[\alpha'_{\textrm{fav}} + \alpha_i + 1, \beta'_{\textrm{fav}} + \beta_i 
+ \delta_i] \,,
\nonumber
\end{eqnarray}
and $B[a,b]$ the Euler $\beta$-function.
The parameters $N_i$, $\alpha_i$, $\beta_i$, $\gamma_i$, and $\delta_i$ come from $D$ FFs at the initial scale and are given in Table III of \cite{deFlorian:2007aj}.
Note that $D^{\pi^+/u}$ in Ref.~\cite{deFlorian:2007aj} differs from $D^{\pi^+/\bar{d}}$.
$J_{\textrm{fav}}$ in~(\ref{e:par_fav}) is similarly defined as $J_{\textrm{fav}}\equiv J_{u+\bar{u}}-J_{\bar{u}}$, where $J_i$ ($i=u+\bar{u},\,\bar{u}$) follows from $I_i$ through $\alpha'_{\textrm{fav}} \to (\alpha_{\textrm{fav}} + 4)$, $\beta'_{\textrm{fav}} \to (\beta_{\textrm{fav}} + 1)$.
The factor $1/(2  I_{\textrm{fav}} J_{\textrm{fav}})$ in~(\ref{e:par_fav}) is convenient and implies $\int_0^1 dz \,z\, H_{(3)}^{\pi^+/u}(z) = N_{\textrm{fav}}$ at the initial scale, 
where $H_{(3)}$ represents the entire second term on the r.h.s.~of~(\ref{e:relation}).
For the disfavored FFs $\hat{H}_{FU}^{\pi^+/(d,\bar{u}),\Im}$ we make an ansatz in full analogy to~(\ref{e:par_fav}), introducing the additional parameters $N_{\textrm{dis}}$, $\alpha_{\textrm{dis}}$, $\alpha'_{\textrm{dis}}$, $\beta_{\textrm{dis}}$, $\beta'_{\textrm{dis}}$.
($I_{\textrm{dis}}$ and $J_{\textrm{dis}}$ are calculated using $D^{\pi^+/d} = D^{\pi^+/\bar{u}}$ from~\cite{deFlorian:2007aj}.)
The $\pi^-$ FFs are then fixed through charge conjugation, and the $\pi^0$ FFs are given by the average of the FFs for $\pi^+$ and $\pi^-$.
The FFs $H^{\pi/q}$ are computed by means of~(\ref{e:relation}).
All parton correlation functions are evaluated at the scale $P_{h\perp}$ with leading order evolution of the collinear functions. 
\begin{figure}[t]
\centering
\includegraphics[scale=0.60]{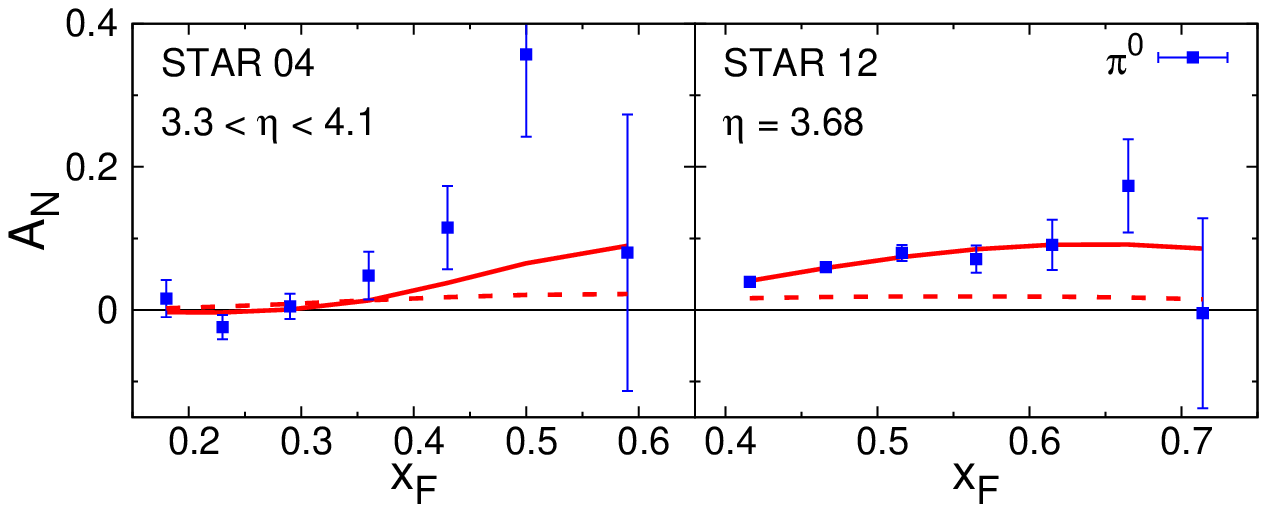} \\
\includegraphics[scale=0.60]{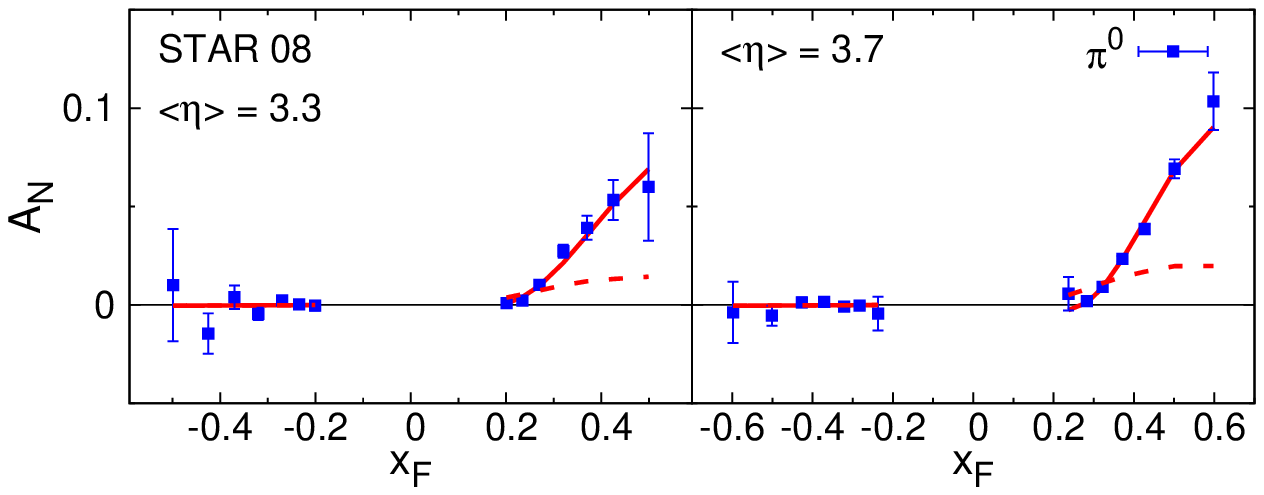} \\
\hspace{0.1cm}\includegraphics[scale=0.60]{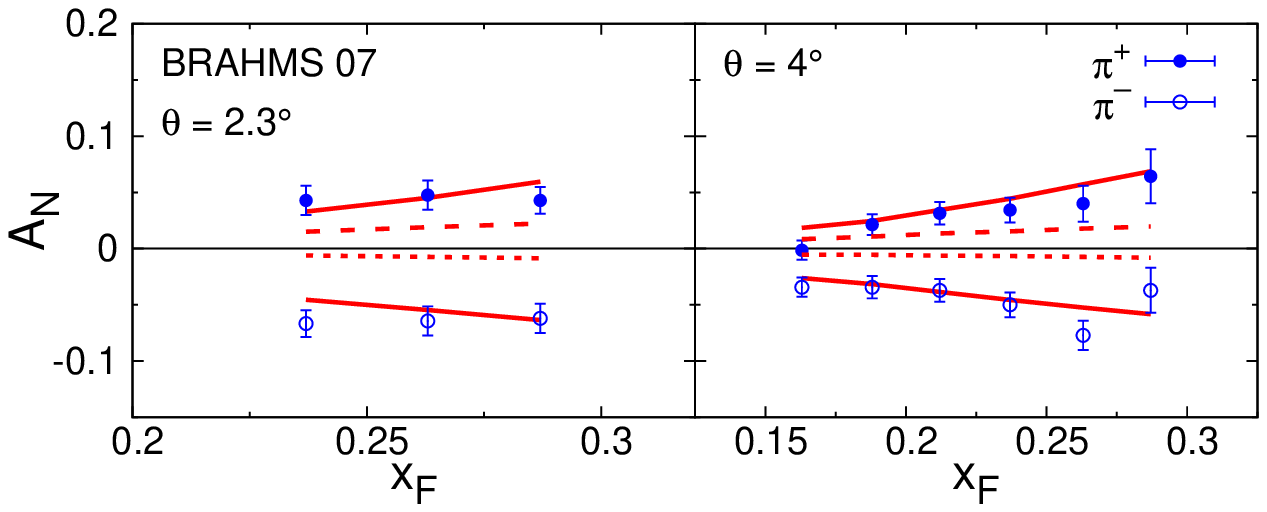}
\vspace{-0.3cm}
\caption{Fit results for $A_N^{\pi^0}$ (data from~\cite{Adams:2003fx,:2008qb,Adamczyk:2012xd}) and $A_N^{\pi^\pm}$ (data from~\cite{Lee:2007zzh}) for the SV1 input. The dashed line (dotted line in the case of $\pi^-$) means $\hat{H}_{FU}^{\Im}$ switched off.}
\vspace{-0.3cm}
\label{f:fit}
\end{figure}

Using the MINUIT package we fit the fragmentation contribution to data for $A_N^{\pi^0}$~\cite{Adams:2003fx,:2008qb,Adamczyk:2012xd} and $A_N^{\pi^\pm}$~\cite{Lee:2007zzh}.
To facilitate the fit we only keep 7 parameters in $\hat{H}_{FU}^{\pi^+/q,\Im}$ free. 
We also allow the $\beta$-parameters $\beta_u^T=\beta_d^T$ of the transversity to vary within the error range given in~\cite{Anselmino:2013vqa}.
All integrations are done using the Gauss-Legendre method with 250 steps.
\begin{figure}[t]
\centering
\includegraphics[scale=0.60]{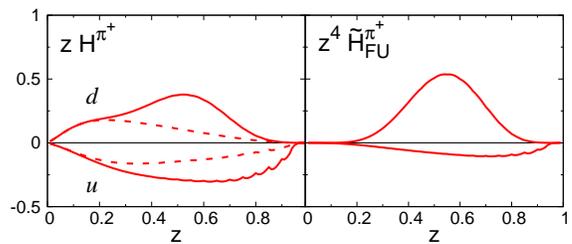}
\caption{Results for the FFs $H^{\pi^+/q}$ and $\tilde{H}_{FU}^{\pi^+/q}$ (defined in the text) for the SV1 input. Also shown is $H^{\pi^+/q}$ without the contribution from $\hat{H}_{FU}^{\Im}$ (dashed line).}
\label{f:functions}
\end{figure}
\begin{figure}[t]
\centering
\includegraphics[scale=0.60]{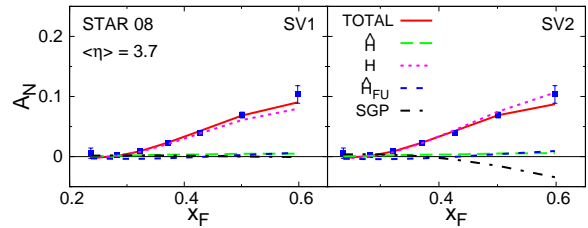}
\vspace{-0.3cm}
\caption{Individual contributions to $A_N^{\pi^0}$ (data from~\cite{:2008qb}) for SV1 and SV2 inputs.}
\vspace{-0.3cm}
\label{f:contrib}
\end{figure}
For the SV1 input the result of our 8-parameter fit is shown in Tab.~\ref{t:fitpar}.
Note that the values for $\beta'_{\textrm{fav}} = \beta'_{\textrm{dis}}$ and $\beta_{\textrm{fav}}$ are at their lower limits, which we introduce to guarantee a finite integration upon $z_1$ in~(\ref{e:sigma_frag}) and a proper behavior of $A_N$ at large $x_F$, respectively.
For the SV2 input the values of the fit parameters are similar, with an equally successful fit ($\chi^2/\textrm{d.o.f.} = 1.10$).
\begin{table}[h]
\vspace{-0.35cm}
\centering
\caption{Fit parameters for SV1 input.}
\label{t:fitpar}
\begin{tabular}{l l l l l l}
\hline
\hline
&&$\hspace{1.0cm}\chi^2/\textrm{d.o.f.} = 1.03$&&&\\
\hline
$N_{\textrm{fav}}$ &=& $-0.0338$ & $N_{\textrm{dis}}$ &=& $0.216$ \\
$\alpha_{\textrm{fav}}$ &=& $\alpha'_{\textrm{fav}} = -0.198$ & $\beta_{\textrm{fav}}$  &=& $0.0$ \\
$\beta'_{\textrm{fav}}$ &=& $\beta'_{\textrm{dis}} = -0.180$ & $\alpha_{\textrm{dis}}$ &=& $\alpha'_{\textrm{dis}} = 3.99$ \\
$\beta_{\textrm{dis}}$ &=& $3.34$ & $\beta_u^T$ &=& $\beta_d^T = 1.10$ \\
\hline
\hline
\end{tabular}
\end{table}

\begin{figure}[t]
\centering
\includegraphics[scale=0.7]{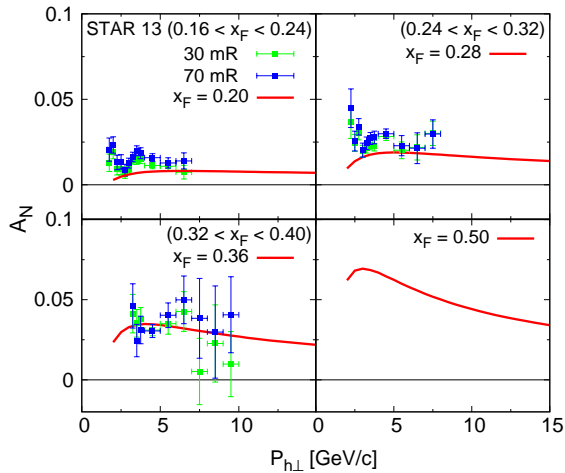}
\vspace{-0.3cm}
\caption{$A_N$ as function of $P_{h\perp}$ for SV1 input at $\sqrt{S} = 500 \, \textrm{GeV}$ (data from~\cite{Heppelmann:2013ewa}).}
\label{f:pt}
\end{figure}

The very good description of $A_N$ is also reflected by Fig.~\ref{f:fit}.
We emphasize that such a positive outcome is non-trivial if one keeps in mind the constraint in~(\ref{e:relation}) and the 
need to simultaneously fit data for $A_N^{\pi^0}$ and $A_N^{\pi^\pm}$.
Results for the FFs $H^{\pi^+/q}$ and $\tilde{H}_{FU}^{\pi^+/q} \equiv \int_{z}^{\infty} \frac{dz_1} {z_1^2}  \frac{1} {\frac{1} {z}-\frac{1} {z_{1}}} \frac{1} {\xi}\hat{H}_{FU}^{\pi^+/q,\Im}(z,z_{1})$ are displayed in Fig.~\ref{f:functions}.
In either case the favored and disfavored FFs have opposite signs.
This is like for $H_1^\perp$ where such reversed signs are actually ``preferred" by the Sch\"afer-Teryaev (ST) sum rule $\sum_h \sum_{S_h} \int_0^1 dz \,z\, M_h \hat{H}^{h/q}(z) = 0$~\cite{Schafer:1999kn}.
Note that the ST sum rule, in combination with~(\ref{e:relation}), implies a constraint on a certain linear combination of $H^{h/q}$ and (an integral of) $\hat{H}_{FU}^{h/q,\Im}$.
In view of that, reversed signs between favored and disfavored FFs like in Fig.~\ref{f:functions} are actually beneficial.
Also depicted in Fig.~\ref{f:functions} is $H^{\pi^+/q}$ when $\hat{H}_{FU}^{\pi^+/q,\Im}$ is switched off. 
As shown in Fig.~\ref{f:fit}, in such a scenario, i.e., by turning off the 3-parton FF, one cannot describe the data for $A_N$.  
According to Fig.~\ref{f:contrib}, the $\hat{H}$ term (including its derivative) in~(\ref{e:sigma_frag}) contributes only very little to $A_N$.
Also the SGP pole term is small, except for the SV2 input at large $x_F$, where its contribution is opposite to the data.
Clearly $A_N$ is governed by the $H$-term in~(\ref{e:sigma_frag}).
This result can mainly be traced back to the hard scattering coefficients:~e.g., for the dominant $qg \to qg$ channel one has $S_H \propto 1/\hat{t}^3$, but $S_{\hat{H}} \propto 1/\hat{t}^2$~\cite{Metz:2012ct} in the forward region where $\hat{t}$ is small.
Finally, Fig.~\ref{f:pt} shows the $P_{h\perp}$-dependence of $A_N$ for $\sqrt{S} = 500 \, \textrm{GeV}$.
Preliminary data from STAR, extending to almost $P_{h\perp} = 10 \, \textrm{GeV}$, show that $A_N$ is rather flat~\cite{Heppelmann:2013ewa}.
The twist-3 calculation agrees with that trend, and also the magnitude of $A_N$ is in line with the data.
Note that the data of Ref.~\cite{Heppelmann:2013ewa} were not included in our fit and that only statistical errors are shown in Fig.~\ref{f:pt}~\cite{Heppelmann:2013ewa}.

%
%
\vspace{0.1cm}
{\it Conclusions}
$\phantom{A}$Collinear twist-3 QCD factorization can be considered the most natural and rigorous approach to the transverse SSA $A_N$ in $p^{\uparrow} p \to h X$.
However, the sign-mismatch issue of the Sivers effect had put this framework into question~\cite{Kang:2011hk}.
Here we have demonstrated for the first time that, despite the sign-mismatch problem, twist-3 factorization actually can describe high-energy RHIC data for $A_N^\pi$ very well if one takes the fragmentation contribution into account.
We re-emphasize that this result is non-trivial.
Since in the twist-3 approach part of $A_N$ can be fixed by spin/azimuthal asymmetries in SIDIS and in $e^+ e^- \to h_1 h_2 X$, we have shown that at present a simultaneous description of all those observables is possible.
We repeat that the fragmentation contribution in twist-3 factorization goes beyond the pure Collins effect.
Independent information on the FFs $H^{\pi/q}$, $\hat{H}_{FU}^{\pi/q,\Im}$ from other sources is needed before one can ultimately claim the intriguing data on $A_N^\pi$ is fully understood.
However, the fact that $\hat{H}_{FU}^{\pi/q,\Im}$ gives a reasonable contribution to (the numerically dominant) $H^{\pi/q}$ (see Fig.~\ref{f:functions}) allows one to be optimistic in this regard.

%
%
\vspace{0.2cm}
\begin{acknowledgments}
We thank S.~Heppelmann, E.-C.~Aschenauer, and R. Fatemi for their effort in making the data of Ref.~\cite{Heppelmann:2013ewa} available to us.
We also thank S.~Scopetta, W.~Vogelsang, and S.~Yoshida for support with regards to the evolution of the transversity distribution.
This work has been supported by the Grant-in-Aid for Scientific Research from the Japanese Society of Promotion of Science under Contract Nos.~24.6959 (K.K.), 23540292 and 26287040 (Y.K.), the National Science Foundation under Contract No.~PHY-1205942 (A.M.), and the RIKEN BNL Research Center (D.P.).
\end{acknowledgments}

\end{document}